\documentclass{article}

\usepackage{authblk} 
\usepackage{graphics} 
\usepackage{epsfig} 
\usepackage{newtxtext}
\usepackage{newtxmath}
\usepackage{url}
\usepackage{mathtools}
\usepackage{soul}
\usepackage{xcolor}
\usepackage{enumerate}
\usepackage{afterpage}
\usepackage{pdfpages}
\usepackage{subfiles} 
\usepackage[numbers]{natbib}

\usepackage{setspace}
\usepackage[letterpaper, margin=1in]{geometry}

\newcommand{\Cb}{{\emph{Cynopterus brachyotis}}}

\begin{document}

\title{Upstroke wing clapping in bats and bat-inspired robots improves both lift generation and power economy}

\author[a,b,1]{Xiaozhou Fan}
\author[b]{Alberto Bortoni} 
\author[a]{Siyang Hao}
\author[b,a]{Sharon Swartz}
\author[a,b]{Kenneth Breuer}

\affil[a]{Center of Fluid Mechanics, School of Engineering, Brown University}
\affil[b]{Department of Ecology, Evolution, and Organismal Biology, Brown University}

\maketitle



\begin{abstract}
Wing articulation is critical for efficient flight of bird- and bat-sized animals. Inspired by the flight of  \Cb, the lesser short-nosed fruit bat, we built a two-degree-of-freedom flapping wing platform with variable wing folding capability. In late upstroke, 
the wings "clap" and produce an air jet that significantly increases lift production, with a positive peak matched to that produced in downstroke. Though ventral clapping has been observed in avian flight, potential aerodynamic benefit of this behavior has yet to be rigorously assessed. We used multiple approaches -- quasi-steady modeling, direct force/power measurement, and PIV experiments in a wind tunnel -- to understand critical aspects of lift/power variation in relation to wing folding magnitude over Strouhal numbers between $St = 0.2 - 0.4$.
While lift increases monotonically with folding amplitude in that range, power economy (ratio of lift/power) is more nuanced. At $St = 0.2 - 0.3$, it increase with wing folding amplitude monotonically. 
At $St = 0.3 - 0.4$, it features two maxima -- one at medium folding amplitude ($\sim 30^\circ$), and the other at maximum folding. These findings illuminate two strategies available to flapping wing animals and robots -- symmetry-breaking lift augmentation and appendage-based jet propulsion. 

\end{abstract}


\footnote{Current Caltech GALCIT postdoc, xzfan@caltech.edu}

\section{Introduction}

\begin{figure}[t]
\centering
\includegraphics[width=3.4in]{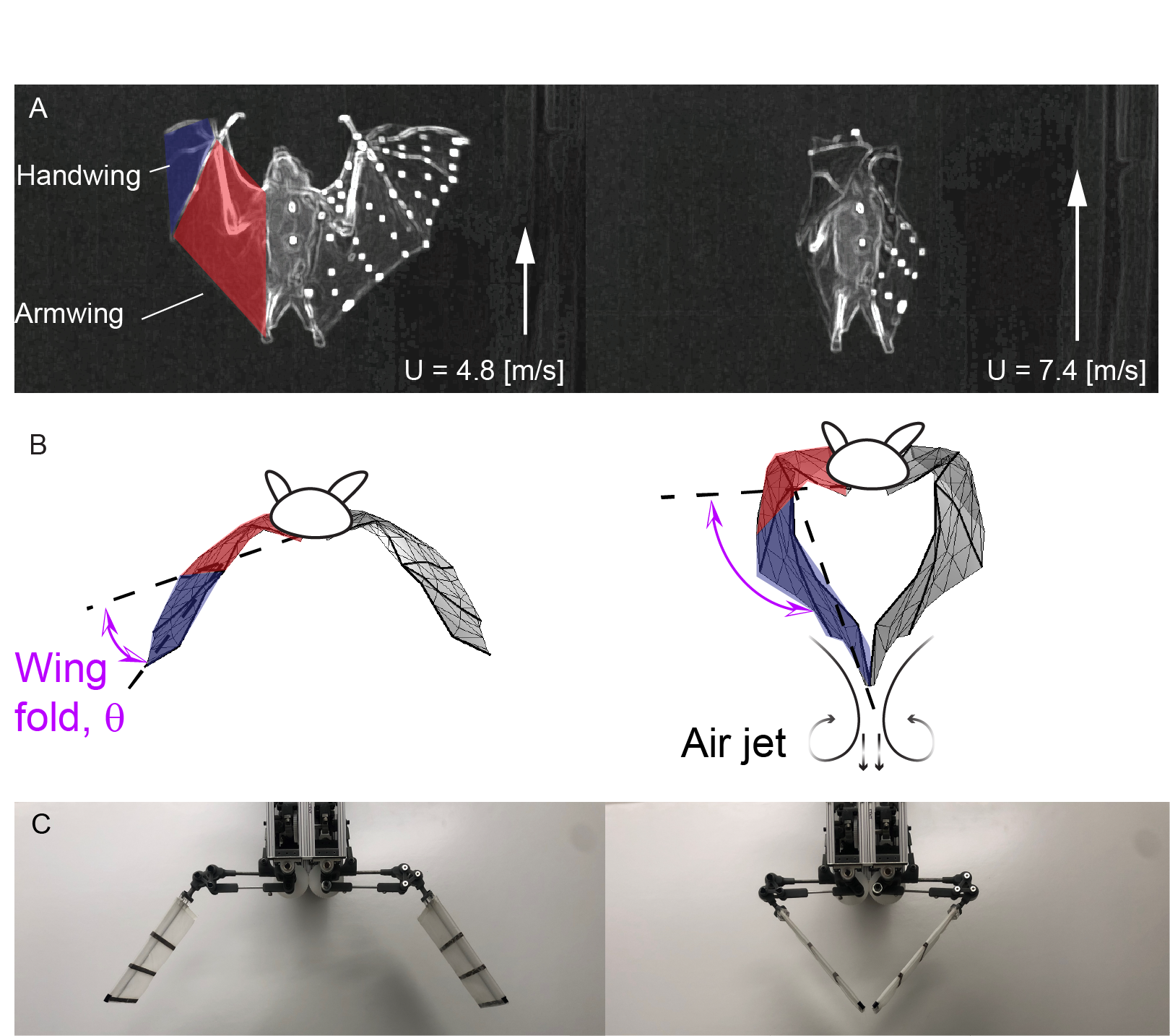}
\caption{(A) Ventral or inferior view of \Cb~in straight and level flight at the start of upstroke (Riskin et al. \cite{Riskin2008}, and videos or raw dataset provided by Fan et al. \cite{Fan2022}) at two different speeds. White markers on the wings are used for motion tracking. The wing folding angle $\theta_o$ at $U = 7.4$ m/s is clearly larger than that at $U = 4.8$ m/s, where the two handwings clap together as a result of large wing folding angles at higher flight speed. 
(B) Frontal views of the meshed bat wings based on motion tracking and stereo-triangulation \cite{Riskin2008,Fan2022}. For flight speed of $U = 4.8$ m/s, at the moment of the maximum wing folding the wingtips remain widely spaced, but for $U = 7.4$ m/s, the two wingtips approach and clap, producing a downward-directed air jet, which generates additional lift. The dashed lines here represent the angle between armwing and handwing as defined in \citep{fan2021a}.
(C) Photo of the two degree-of-freedom, bat-inspired robot, ``Flapperoo'', capable of performing independent flapping and folding motion.}
\label{fig:intro}
\end{figure}

Bats fly with highly-articulated wings; in particular, the kinematics of the handwing, from the wrist to the wing tip, vary significantly with flight speed \cite{Muijres2008,Hubel2010,Hubel2016,fan2021a,Fan2022}. During upstroke, in addition to wing elevation, bats' wrists flex, which rotates the handwing with respect to the armwing about a chord-wise axis through the wrist, and fold the wings \cite{Hubel2016} (Fig.~\ref{fig:intro}A). It has been suggested that folding reduces the inertial power of wing elevation \cite{Riskin2012}, and that the reduced wing wetted surface area due to folding leads to a decrease in negative lift \cite{Hubel2010,Hubel2016}. During late downstroke, due to wing folding, the additional angular velocity of the handwing increases its effective angle of attack and effective velocity, which contribute to greater lift \cite{Sekhar2018,Fan2021b}. It is therefore no surprise that bio-inspired robotic flyers with wing folding capability also demonstrate superior performance, such as flight endurance \cite{send2012,Wissa2012}, compared to those that flap but do not fold \cite{Bie2021}. \footnote{Note that in this study, we employ a specific definition of wing folding: the additional rotation of the outboard wing section (handwing) relative to inboard wing section (armwing) in the stroke plane (Fig.~\ref{fig:intro}B) \cite{send2012,Wissa2012}. This differs from the planar protraction/retraction of wings along the wingspan \cite{Vejdani2019}, which has also been referred to as wing folding \cite{Bahlman2013,Bahlman2014,Stowers2015}.} 

In some cases, bats fold their wings so much during upstroke \cite{Riskin2008}, and the two wingtips touch and clap (Fig.~\ref{fig:intro}B). This phenomenon occurs not only in multiple bat species \cite{Gould1988,Boonman2014}, but also in small birds that hover, such as Warbling White-eyes (\textit{Zosterops japonicus}
), and Gouldian Finches (\textit{Erythrura gouldiae}
) \cite{Chang2011,Chang2013}. 


In this manuscript, we report on efforts to understand how wing folding may influence flight performance in bats in terms of aerodynamics and energetics. This is difficult or impossible to ascertain in live animals, and thus we designed, built and tested a mechanical robot, ``Flapperoo'', that abstracts key features of the bats' wingbeat kinematics, allowing us to manipulate relevant variables.

By combining direct force/power measurements, time-resolved particle image velocimetry (PIV), and quasi-steady modeling, we explain the measured forces/power using validated blade-element momentum theories using the observed kinematics \cite{fan2021a,Fan2022}. Additionally, the measured forces, power and PIV help to determine the effectiveness and limitations of the reduced-order computational model. If folding enhances flight performance of membrane wings, an exploration of this enhancement using robotic platform is a crucial first step in the design process, which identifies how best to incorporate wing folding into a flapping wing engineered device.

However, there are as yet no systematic studies on the unique wing folding/clapping phenomenon. Few robotic studies have implemented the actively-actuated wing folding motion observed in bat and bird species  \cite{send2012,Chen2021} and, to the best of our knowledge, none have reported the power and aerodynamics of varying wing fold amplitudes. Both Chen et al. \cite{Chen2021} and Kashi et al. \cite{Kashi2020} built robotic flappers to replicate folding in wings or during flipper clapping, but no lift benefits were observed.

\begin{figure}
\centering \includegraphics[width=5in]{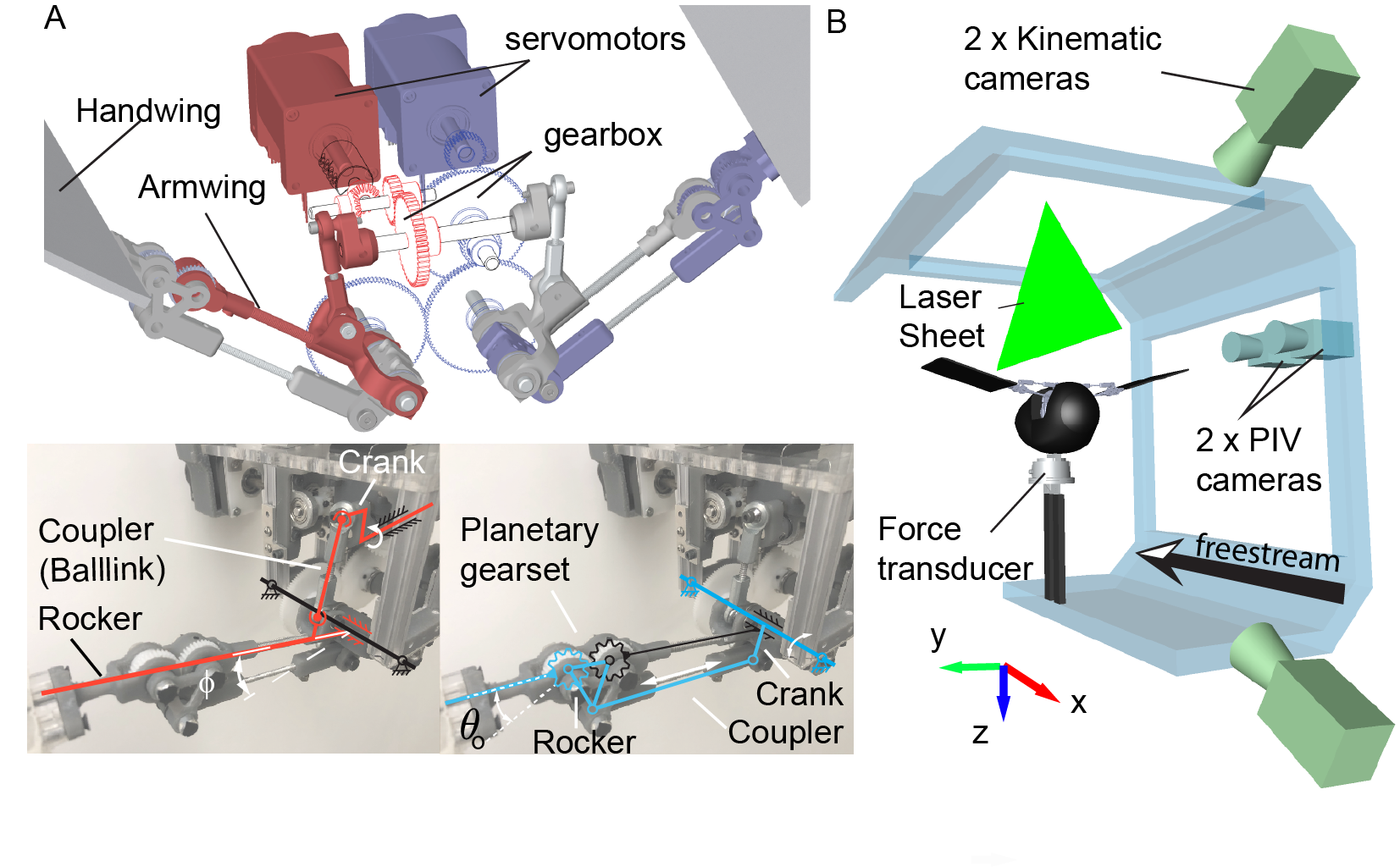}
\caption{Model design and experimental setup. (A) Flapperoo performs independent wing flapping (mechanism in red, with fixed magnitude $\phi$) and folding (mechanism in blue, with magnitude denoted as $\theta_o$) motions, which are realized by two four-bar linkages, each driven by a servomotor. The left and right wing motions are symmetric. (B) Experimental setup. Flapperoo is mounted on a six-axis force/torque transducer in the wind tunnel. The freestream points to the negative of the $x$-axis, and the $z$-axis is direction of gravity. The model is mounted upside down (ventral-side up) for unobstructed access to PIV laser sheet. }
\label{fig:exprSetup}
\end{figure}

To assess the role of wing folding in flapping flight, both flapping and folding motions are programmable for Flapperoo. The bio-inspired robot allows direct measurement of force and power in a manner that is not possible from animals. 
The wing movement mechanism is composed of two four-bar linkages, driven by two servomotors, in which one controls wing flapping, a movement of the ``armwing'', and the other controls wing folding, a movement of the ``handwing'' relative to the armwing (see Fig.~\ref{fig:intro}C, Fig.~\ref{fig:exprSetup}A, and refer to Sec.~\ref{sec:MM} for more detailed description). 

We define up- and downstroke  by the motions of the armwing and set them to be equal in duration. Studies of hovering Japanese White-eyes, \textit{Zosterops japonicus}, and Gouldian Finches, \textit{Erythrura gouldiae}, have classified wing clapping as part of downstroke \cite{Chang2011,Chang2013}, based on the trajectories the wing tips. However, the primary flight motor of vertebrates is the pectoralis major muscle \cite{vaughan1959}, and for this reason, we deem the motion of armwing to be a more accurate reflection of up- vs. downstroke. Additionally, the cyclic motion of the armwing is more stable and consistent than that of the deforming handwing. Using this framework, bat wing clapping occurs after the armwing has reached the bottom of the downstroke and begun upstroke, although the handwings, which perform the clap, are still moving downward (ventrally) at this point in the wingbeat cycle.
For all the wing folding cases we considered, folding begins in the late downstroke. 

We carried out tests over a range of freestream velocities and wing folding amplitudes in a wind tunnel (Fig.~\ref{fig:exprSetup}B). This corresponded to Strouhal numbers, $St$,  (defined as $St = fA/U$, where $f$ is the flapping frequency, $A$ is the vertical wingtip displacement when there is no folding, and $U$ is the freestream velocity \cite{Taylor2003,Riskin2008}) over the range of $ 0.2 - 0.4$ -  a particularly relevant range for flying animals \cite{Taylor2003b}. We used the measured robot kinematics as input to a quasi-steady computational model \cite{fan2021a,Fan2021b,Fan2022}, which estimated aerodynamic force and power. Then, we compared force and power, estimated from the computational simulations, with the directly measured forces (from a force transducer) and power consumption (from the motor's driving current and voltage). To quantitatively characterize the air jet produced, we measured the flow in a vertical (parasaggital) plane between the two wings (Fig.~\ref{fig:exprSetup}B) using time-resolved, particle image velocimetry (PIV), and performed a Reynolds-averaged control volume analysis to assess the forces associated with this flow \cite{Bohl2009,Mathai2023}. 

\section{Materials \& Methods} \label{sec:MM}

We designed and built a two-degree-of-freedom (DoF) flapping wing robot, Flapperoo, capable of both wing flapping and folding.
The flapping motion (red, Fig.~\ref{fig:exprSetup}A) is realized by a four-bar linkage mechanism. A crank rotates continuously in one direction to pull/push the ball link (coupler), causing the armwing (rocker) to oscillate up and down; the angle between the rocker and the horizontal plane is denoted as the flapping angle $\phi$, which is fixed. 
To fold the handwing (blue, Fig.~\ref{fig:exprSetup}A), an independent crank is driven reciprocally, dragging the coupler to move in and out. This motion pivots the spur gear (planet gear, blue) on the rocker (carrier) to rotate against the fixed gear mounted on the armwing (sun gear, black), so that the folding angle is amplified ($2:1$ ratio). The handwing is fixed to the spur gear, and the angle it makes relative to the armwing (black) is the wing folding angle, denoted as $\theta$. The folding of wings begin at $t/T = 1/3$ of a cycle, in the downstroke (black arrow in Fig.~\ref{fig:temporalAndScalingPIV}A2), and is sinusoidal. A streamlined body for Flapperoo was hand-sculpted using foam and houses the flapping mechanism (Fig.~\ref{fig:exprSetup}B). The body and wings were painted matte black to reduce light scattering during PIV experiments.

The wing ribs were laser-cut from flat balsa wood strips with a chord length of $c = 200$mm, and a thickness of $5$mm. The equally-spaced ribs were glued onto a $200$mm carbon-rod spar before being covered with rip-stop, kite-cloth fabric. The handwing and armwing are equal in length. The wing assembly weighs $10$-grams. \textcolor{blue}{To isolate the effect of handwing structure, the $100$mm armwing skeletal frame is not covered with fabric.}

The flapping and folding movements are actuated by two brushless, rotary servomotors with integrated encoders (BE163CJ-NFON, Parker Hannifin Corp., Rohnert Park, CA), controlled by a servo controller (DMC-4060, Galil Motion Control, Rocklin, CA). The motors are controlled using custom written code in software developed for the controller (GDK, Galil Motion Control), and operate in ‘PVT’ mode, where the user defines a list of target positions and velocities at a series of defined times. The controller then moves the motor through a profile that reaches each target position at the target time, moving at the target velocity. The list of target values was calculated using a Matlab script. 

The flapping motor was programmed to execute a constant speed rotation.  We used a testing frequency of 3 Hz  for all investigations reported here. The folding motor was programmed to start the fold motion at $t/T = 1/3$ and to end at $t/T = 1$.  Due to the coaxial nature of the rocker during flapping and the crank during folding, friction will cause one to rotate with the other, and thus, even for the no folding-case ($\theta_o = 0$), we drive folding at a low level to counter this friction at a non-zero amplitude (Fig.~\ref{fig:temporalAndScalingPIV}A1). This small base offset is added to all other cases. 

The robot was tested in a closed-loop wind tunnel at Brown University \cite{Breuer2022} (Fig.~\ref{fig:exprSetup}B). The test section is $1.2 \times 1.2$ meters in cross section and 4 meters long.  We varied the freestream velocity to achieve the desired Strouhal number, $St = fA/U$, where $A$ is the vertical wingtip displacement when there is no folding ($\theta_o = 0$), and $U$ is the freestream velocity for the fixed 3 Hz testing frequency \cite{Taylor2003,Riskin2008}. We introduce an additional, local Strouhal number to describe jet scaling during wing clapping, $St_{clap} = f \theta_o b/U$, where $b$ is the length of each individual handwing.

The robot was mounted on a six-axis force/torque sensor (Gamma IP65, ATI Industrial Automation, NC) and force/torque measurements were recorded using an A/D converter (USB-6343, National Instrument, TX) at $1000$Hz. Angular velocity and torque were recorded at $512$Hz from the motors using the Galil motion controller, and their inner product was computed to yield mechanical power.
The net lift was normalized by the dynamic pressure of the freestream and the wing area: $1/2 \rho U^2 2*b c$.
The net power was similarly normalized by the dynamic pressure of the freestream and the wing area and chord as: $1/2 \rho U^3 2*b c$.

For each trial, we first acquired data with the handwings attached, and then removed the handwings and repeated each measurement with the same set of parameters. The trials without handwings record the inertial forces and torques associated with motion of the mechanical components (e.g., gears, cranks, etc). \textcolor{blue}{Note that the inertia power from handwing ($\sim 9[g]$) alone is negligible, and is thus not considered}. A Butterworth low-pass filter was applied to these data with a cut-off frequency $f_c = 5f (15 Hz)$. The net aerodynamic force or power is the difference between these two conditions (see also Bahlman et al. \cite{Bahlman2014}).

White reflective markers, placed at the leading edge of the wings, were tracked from video acquired by two Phantom Miro 340 cameras (Vision Research Inc.) at $800$fps. The video and force data acquisition were synchronized using an Arduino microcontroller.
The cameras were calibrated and and the video digitized using DLTdv8 \cite{Hedrick2008}. Subsequently, the  wing kinematics were used in a quasi-steady flapping flight model to predict aerodynamic force and power \cite{Vejdani2019,fan2021a,Fan2021b,Fan2022}. 

The particle image velocimetry (PIV) experiment used a Nd:YLF double-pulsed laser (DM30, Photonics Industries, Ronkonkoma, NY), employed at $500$Hz with an energy output of approximately $30$ mJ/pulse. The vertical laser sheet was aligned with the flow direction and passed through the Flapperoo's body midline, the center of the gap between the wingtips (Fig.~\ref{fig:exprSetup}B). The test-section was seeded with neutrally buoyant, helium-filled soap bubbles (diameter 0.3 mm), released upstream of Flapperoo (Lavision Inc., Germany). The bubbles illuminated by the laser sheet were imaged by two high-speed cameras (Photron NOVA R2, 2048 x 2048 pixels), positioned side by side with overlapping fields of view, producing a combined field of view of $600 \times 400$mm. We used DaVis PIV software v10 (LaVision Inc., Germany) to perform image cross-correlation, and MATLAB code to obtain time-resolved phase-averaged velocity fields in $x$ and $z$ ($N = 83$ bins/cycle, where each bin contained $40$ repetitions \cite{Onoue2016}).



The aerodynamic forces that result from wing clapping, $\vec{L}$, were derived from the Navier-Stokes equation using the control volume (C.V.) analysis approach (Fig.~\ref{fig:temporalAndScalingPIV}C, see Bohl \textit{et al.} \cite{Bohl2009} for a similar but simpler analysis):
\begin{multline}
\vec{L} = \underbrace{\frac{\partial}{\partial t} \int_{C . V .} \rho \vec{v} dV}_\text{\clap{Acceleration~}}
+ \underbrace{\int_{\text {C.S. }} \rho \vec{v}(\vec{v} \cdot \vec{n}) d A}_\text{\clap{Momentum flux~}}
- \underbrace{\int_{\text {C.S. }}-p \cdot \vec{n} d A}_\text{\clap{Pressure~}}, 
        \label{eq:CV_final}
\end{multline}
where $\rho$ is the air density, $\vec{v}$ the velocity field of the flow, $\vec{n}$ the surface normal of each control surface (C.S.), and $p$ the pressure acting on each surface. 


All derivatives were evaluated using a Savitzky–Golay filter (order 2, frame length 5) \cite{Savitzky1964}. The first term is a volume integral of the acceleration field along the vertical $z-$ direction, the second term is the transport of $z-$axis momentum across each downstream boundary surfaces of the control volume, and the third term is a line integral that gives the pressure difference between bottom ($3$) and top ($4$) surface in Fig.~\ref{fig:temporalAndScalingPIV}C. 

To obtain the clapping lift coefficient $C_{L,clap}$, we assume that the width of the jet is uniform into the depth of the measurement plane ($y$-axis in Fig.~\ref{fig:exprSetup}B) around the time of wing clapping, and that the width is of the order of twice the length of the handwing, $2b$, consistent with PIV studies of wing clapping in birds \cite{Chang2011,Chang2013}. Normalized by the dynamic pressure, $q$, and surface area of the whole wing, $S = 2bc$, the lift coefficient from the PIV experiment is thus $C_{L} = L 2b/qS = L 2b/ q2bc = L/qc$.

\section{Results}

\begin{figure}
\centering
\includegraphics[width=1.5in]{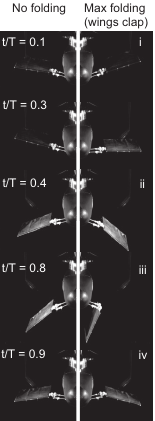}
\caption{Frontal view of wing kinematics in the wind tunnel at five key snapshots of one wingbeat cycle. Left and right panels are for cases without folding ($\theta_o = 0$) and maximum folding ($\theta_o = 100^{\circ}$, wings clap) respectively.}
\label{fig:kineSnaps}
\end{figure}

\begin{figure}
\centering
\includegraphics[width=5in]{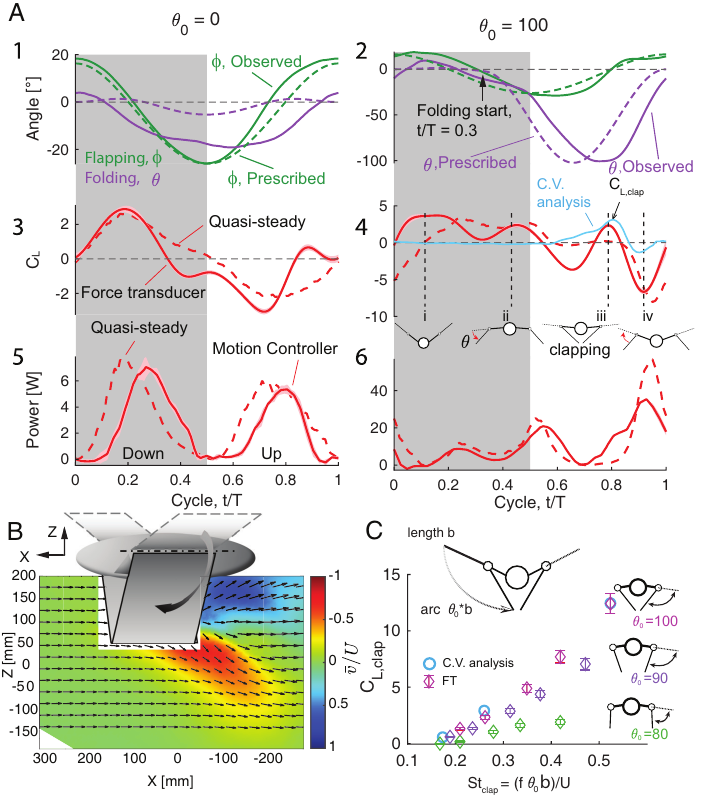}
\caption{
(A) Kinematics, lift coefficient, and power consumption over the wingbeat cycle for the near-zero ($\theta_o = 0$) and maximum folding angle cases ($\theta_o = 100^{\circ}$, wings clap)
Freestream velocity, $U = 4$ m/s. Solid line: observed left wing kinematics;  dashed line: commanded motion; left and right wing movements are symmetric. 
Grey shading indicates downstroke as defined by armwing motion.  Force coefficients of lift and and power measurements: solid lines indicate average force (40 cycles), shading indicates rms variance. $C_{L,clap}$ is defined as the local peak of the $C_L$ during wing clapping. The control volume analysis, derived from PIV, is presented as a solid blue line. Four snapshots (i, ii, iii and iv) highlight instances of interest in the wingbeat cycle (Fig.~\ref{fig:kineSnaps}).
(B) Phase-averaged (40 cycles) PIV velocity field at time point iii, maximum wing folding during clapping. The instantaneous velocity field is depicted by black arrows, with vertical components normalized by freestream velocity ($U = 4$m/s). 
(C) Scaling of $C_{L,clap}$ as a function of a locally-defined Strouhal number, $St_{clap}$, from  transducer-measured forces (FT) for three folding angles, and from the PIV-derived control analysis calculations. The uncertainty bars are the standard deviation calculated from 40 cycles.
}
\label{fig:temporalAndScalingPIV}
\end{figure}

\subsection{Robot performance}

We present results for two cases at the extreme of the robot's kinematic space: near-zero wing folding, $\theta  = 0^\circ$,  and extreme wing folding with a wing clap $\theta = 100^\circ$ (Fig.~\ref{fig:kineSnaps}). For both cases the freestream $U = 4$m/s, which corresponds to $St = 0.21$.

We observed smooth flapping motion that followed the prescribed motion with reasonable fidelity (Fig.~\ref{fig:temporalAndScalingPIV}A1 and A2, green lines).  
For the near-zero prescribed folding angle case (Fig.~\ref{fig:temporalAndScalingPIV}A1), the folding angle of the handwing varies passively over the wingbeat cycle, with values ranging from $-20^\circ$ to $5^\circ$. The observed folding angle is closer to the prescribed value for the larger folding angle case, with values similar to those prescribed in magnitude but with a phase lag of about $t/T = 0.1$ (Fig.~\ref{fig:temporalAndScalingPIV}A2). \textcolor{blue}{See also the hand wing total angles with respect to mid-stroke plane in supplementary Fig.~\ref{sfig:hdMid}.}

\subsection{Lift}

The mechanics and energetics of the two folding cases differ substantially. The magnitude and trend of the aerodynamic lift coefficient, $C_L$ ($C_L = F_z/0.5\rho U^2 S$, where $F_z$ is the dimensional force,  $\rho$ is the fluid density, and $S$ is the total area of the two handwings), for the case of near-zero folding is similar when measured by the force transducer (FT) and predicted by our quasi-steady model,  \cite{fan2021a,Fan2022} (Fig.~\ref{fig:temporalAndScalingPIV}A3). Positive lift is generated during downstroke, and negative lift during upstroke. The quasi-steady model diverges from the FT data primarily by an overprediction (more positive lift) of $C_L$ in late downstroke/early upstroke ($t/T = 0.3 - 0.6$), and an underprediction (less positive lift) around the upstroke to downstroke reversal ($t/T = 0.8 - 1$). 
For the extreme wing folding case (Fig.~\ref{fig:temporalAndScalingPIV}A4), downstroke is  responsible for most of the positive lift ($t/T = 0 - 0.5$), but there is more variation across the wingbeat cycle and a more pronounced difference between the lift estimates of the measurement and the quasi-steady model . Specifically, the model produces a more pronounced underprediction of $C_L$ in comparison to the FT measurement at the beginning of downstroke (snapshot i), and slightly overpredicts lift around the mid-downstroke (between snapshots i and ii). The quasi-steady and FT measurement agree well late in downstroke (snapshot ii).  At time $t = 0.8$, when the wing claps (Fig.~\ref{fig:temporalAndScalingPIV}A4, snapshot iii), the quasi-steady model fails to predict the peak in lift generation $C_{L,clap}$ in Fig.~\ref{fig:temporalAndScalingPIV}A4. Towards the end of upstroke ($t/T \sim 0.9$, Fig.~\ref{fig:temporalAndScalingPIV}A4, snapshot iv) when the wing has completed the upstroke movement from the ventral to the dorsal side of the body and prepares for the next downstroke, the quasi-steady model captures the negative lift measured by the force transducer, albeit with a slight overprediction and phase delay.

\subsection{Power Consumption}

Computations of the quasi-steady model predict moderate peaks in power consumption during the middle of the downstroke and the upstroke (Fig.~\ref{fig:temporalAndScalingPIV}A5). These predictions are supported by the power measurements from the robot, although there is a delay (with the quasi-steady model leading by $t/T = 0.1$) in the power maximum. When wing folding is at its maximum ($\theta_o = 100$, Fig.~\ref{fig:temporalAndScalingPIV}A6), the model predicts a more complex time course for power consumption, but the overall pattern of power use measured from Flapperoo is a close match to the simulation,  with the exception of an underprediction of power at snapshot iii ($t/T \sim 0.8$), and a sharp rise in predicted power late in the upstroke at $t/T \sim 0.9$,  (snapshot iv) which is not observed by the measurement.

\subsection{Velocity field}

PIV measurements of the flow field along the midline plane of Flapperoo show a prominent jet that is directed downward and to the rear of the ``animal'', which appears as a result of wing clapping at  $t/T \sim 0.8$, (snapshot iii, and  Fig.~\ref{fig:temporalAndScalingPIV}B). Because the velocity measurement is restricted to the centerline plane between the two wings, the control volume (CV) analysis of the aerodynamic forces associated with this jet (Fig.~\ref{fig:temporalAndScalingPIV}A4 - blue line), are only expected to yield accurate force estimates during the portion of the wingbeat cycle in which wing clapping occurs (refer to sec. \ref{sec:MM} for a more detailed description).  During this time period, $C_L$ agrees well with FT measurement. 
During other portions of the wingbeat cycle, flow in this plane closely follows the freestream and  $C_L$ = 0 (see video in supplementary S1). The $C_L$ calculated from the CV approach is independent of the CV boundary (once a large enough volume is chosen). 

The jet is produced as a result of left and right wings clapping, ``squeezing'' air out. We predict that the strength of the jet created by two wings clapping will depend on the distance between the wingtips, which, in turn,  depends on the clapping amplitude, $\theta_o$) and the speed of the approaching wings,  $f \theta_o b$, where $f$ is the flapping frequency and $b$ is the length of the handwing, measured from the wrist to the wingtip.
We define a local lift coefficient, $C_{L,clap}$, from the peak lift force during the  clap (Fig~\ref{fig:temporalAndScalingPIV}A4), and observe a clear dependence of $C_{L,clap}$ on clapping amplitude, $\theta_o$, and the wingtip velocity (normalized by the flight speed to form a clapping Strouhal number: $St_{clap} = f \theta_o b/U$)(Fig~\ref{fig:temporalAndScalingPIV}C). 

\begin{figure}
\centering
\includegraphics[width=5in]{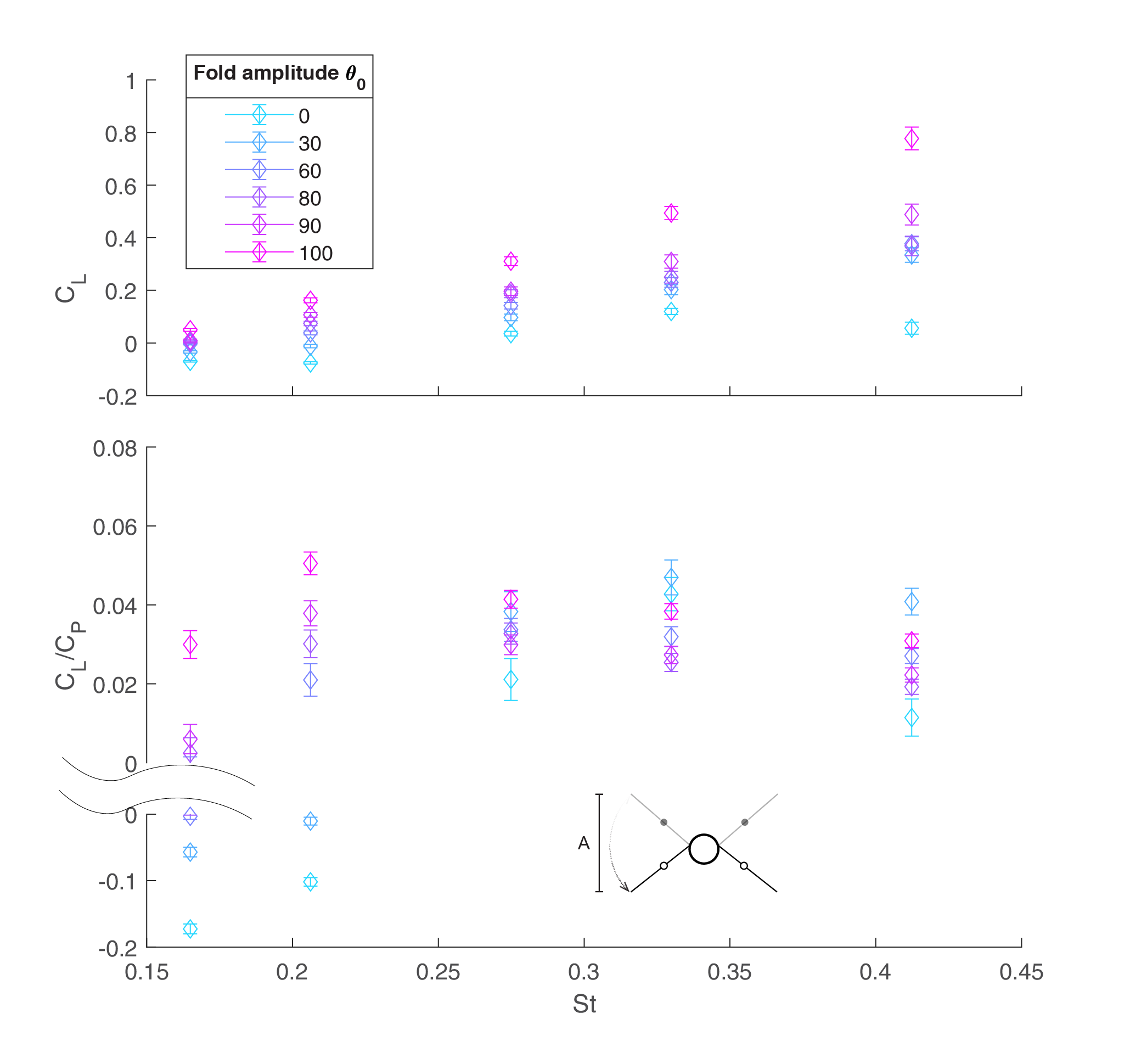}
\caption {Cycle-averaged (A) lift coefficient, and (B) power economy as a function of Strouhal number and maximum folding angle $\theta_o$. The aerodynamics of the body, and the inertia of the mechanism in motion are subtracted.} 
\label{fig:cycAverPwrEco}
\end{figure}

\begin{figure}
\centering
\includegraphics[width=5in]{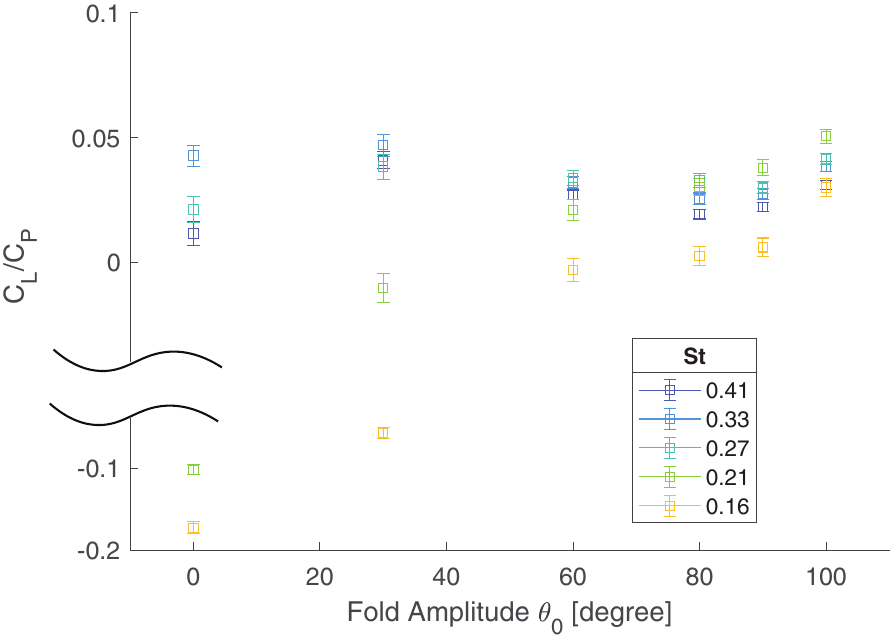}
\caption {Cycle-averaged power economy as a function of folding amplitudes $\theta_o$.} 
\label{fig:PwrEcoVsFA}
\end{figure}

Wing folding confers aerodynamic benefits but requires actuation, and so incurs energetic costs. Examination of the cycle-averaged lift coefficient, $C_L$, and power economy (the ratio of $C_L$ to $C_P$) as a function of Strouhal number, $St$ (Fig.~\ref{fig:cycAverPwrEco}A) shows that for the non-folding case, $\theta_o = 0$, the cycle-averaged $C_L$ is close to zero for all values of $St$ 
(the upstroke-downstroke wing motion is symmetric when wing folding amplitude, $\theta_o$, is non-zero, and hence no net lift is expected). With wing folding, the cycle-averaged lift coefficient increases with $St$, up to a maximum value of $C_L \sim 0.8$ for $(St, \theta_o) = (0.41, 100^\circ)$.  For a given Strouhal number the lift coefficient increases with wing folding amplitude, $\theta_o$, and increases more steeply close to the clapping condition (see supplementary Fig.~\ref{sfig:ClvsFold}, for $\theta_o = 90$ to $100$).

Power economy, $C_L/C_P$, does not monotonically increase with fold angle nor fold angle (Fig.~\ref{fig:cycAverPwrEco}B and Fig.~\ref{fig:PwrEcoVsFA}). For low to moderate $St$ ($St = 0.16 - 0.27$ - moderate to fast flight), power economy increases monotonically with folding amplitude $\theta_o$ and peaks at the highest fold angle $\theta_o = 100$. For higher $St$ ($St = 0.33 - 0.41$ - lower flight speed), the most economical solution to gain lift occurs around $\theta_o = 30$, although Power economy does start to increase again due to wing clapping at high folding amplitude $\theta_o$ (Fig.~\ref{fig:PwrEcoVsFA}). Thus at high $St$, power economy features twin peaks, one at moderate folding, and the other at extreme folding amplitude. Note that right around $St = 0.27$, both folding angle of $\theta_o = 30$ and $100$ are indistinguishably optimal.

\section{Discussion}

Generation of propulsion by jetting for locomotion is well-known in the biological world. Jellyfish contract circular muscle fibers in their hemispherical bell-shaped bodies, shrinking total volume and ejecting water to effect the jet \cite{Costello2020}. Circular muscles around the squid funnel of the squid mantle cavity control intake and propulsive outflow of water \cite{Anderson2000}. The vortex dynamics of these volume reduction-based approaches to jet propulsion have received considerable attention in the fluid dynamics community. Similarly, appendages can generate jet propulsion.  Sea Lions clap their rear flippers during swimming \cite{Perrotta2022}, and wing clapping has been observed in small birds during hovering \cite{Chang2011} and by some bats during fast flight \cite{Riskin2008,Fan2022}. 

In addition to  \Cb~, the inspiration for this study, biologists have long observed wing clapping behavior in bat species from the family Pteropodidae: \textit{Eonycteris spelaea, Macroglossus sobrinus, Eidolon helvum,} and species in the genus \textit{Rousettus} \cite{Gould1988,Boonman2014}. The foci of these reports, however, were the clicking sounds generated by wing clapping as a form of biosonar \cite{Gould1988,Boonman2014} and not any potential aerodynamic effects. 

A robotic platform that mimics sea lion flippers demonstrated clapping-based jet propulsion but did not show force augmentation associated with the jet \cite{Kashi2020}. In contrast, we find that lift is significantly enhanced by wing clapping, with clapping generating a force close to that generated during the downstroke (snapshot iii in Fig.~\ref{fig:temporalAndScalingPIV}A4, and Fig.~\ref{fig:kineSnaps}). This enhancement comes primarily during the upstroke, not generally considered a lift-generating phase of the wingbeat cycle \cite{Shyy2013, Hubel2010, Hedenstrom2007}.

\subsection{Force estimates from PIV}

To calculate the lift generated by wing clapping by hovering birds, Chang et al. \cite{Chang2011,Chang2013} conducted frontal and parasagittal plane PIV measurements during hovering flight of these birds, and a coherent vortex ring is visible as a result of ventral wing clapping. Lift is estimated from the circulation of this vortex ring during downstroke, under the assumption of an aerodynamically passive upstroke, and explains approximately $80\%$ of weight support. Here, we do not assume a vortex ring structure, but apply a control volume (CV) approach,  which 
lends itself well to the spatially and temporally complex flow in terms of pressure, momentum flux, and acceleration of fluid by the wing motion. 

By phase-averaging the flow field over $40$ wingbeat cycles, we found that the contribution from turbulent velocity fluctuations (see Supp. Mat.) is negligible in estimating $C_{L,clap}$ compared to the mean (phase-averaged) values, suggesting that a coherent vortex ring dominates lift production. This agrees with the observation by Chang et al.\cite{Chang2011,Chang2013}.

Net pressure and fluid acceleration start to build well before the wings clap (Fig.~\ref{fig:temporalJetComp}), suggesting that the wing-wing interaction occurs much earlier than the actual instant of clapping (see also \cite{Lehmann2015}), and lasts a substantial fraction of a wingbeat ($t/T = 0.65 - 0.8$ in Fig.~\ref{fig:temporalJetComp}). Indeed, the wings do not need to be physically touching to effect the jet, but instead can be separated by a small distance (Fig.~\ref{fig:temporalAndScalingPIV}C). 

\subsection{Effectiveness and limitations of the quasi-steady model}
\label{discussion:QS effectiveness}

For the case without wing folding (near-zero case), the agreement between the force measurements and the quasi-steady predictions is generally very good, both in terms of the overall trend and the magnitude of the lift force (Fig. \ref{fig:temporalAndScalingPIV}A3).  Because the the wing motion -- up- and downstroke -- is symmetric (Fig. \ref{fig:temporalAndScalingPIV}A1), the model predicts the upstroke and downstroke forces to be equal and opposite, and the measurements largely support this -- with the deviation due to slight asymmetry in the realized kinematics. Measured and predicted forces differ just ahead of the down- and upstroke reversals, at $t = 0.4$ and $t = 0.9$ respectively. At mid downstroke and upstroke ($t/T \sim 0.25$ and $\sim 0.75$), the velocity and effective angle of attack (eAoA) peak (supplementary Fig.~\ref{sfig:eAoA}A). As a result, the lift force peaks at these points in the cycle. The wing slows as it reaches stroke reversal, and the lift force predicted by the quasi-steady analysis approach zero.  However, the measured lift drops to zero somewhat earlier in the wingbeat cycle.  

This discrepancy between measured and predicted lift exposes one of the key weaknesses of the quasi-steady approach: it fails to account for the wake capture around the up- to downstroke reversal. At the end of downstroke, a leading edge vortex (LEV) forms on the dorsal side of the wing, 
\cite{Sane2002,Windes2018,Fan2021}. This vortex contributes additional lift and, as long as it stays attached, can be modelled accurately by the quasi-steady approach using a generalized lift coefficient \cite{Dickinson1999}. However, for large effective angles of attack, the vortex separates and when it moves away from the wing, the lift force drops precipitously. Simulations of vortex formation and shedding of a 2D airfoil have shown that for eAoA = 72$^\circ$, the vortex is much stronger and sheds more quickly than for the case of eAoA = 4.5$^\circ$, in which the core is weaker and remains attached \cite{Wang2000}. Here, the eAoA at the 3/4-span section for the zero folding case ($\theta_o = 0$) varies between -40 to 40 degrees (SFig.~\ref{sfig:eAoA}A). We suggest that at these points in the wingbeat cycle the LEV is likely shed into the wake, leading to the early drop in the lift coefficient, which is not captured by the low-order model.

For the case of 100$^\circ$ wing folding, the amplitudes of the forces are accurately captured and the overall trends match well, but we do see significant differences between measurements and the model predictions (Fig. \ref{fig:temporalAndScalingPIV}A4). In particular, at the start of the downstroke ($t/T \sim 0.1$), at the beginning of the upstroke ($t/T \sim 0.6$), and lastly mid-upstroke ($t/T \sim 0.8$).  The first two of these points in the cycle can be explained using the same reasoning as with the straight wing.  In this case, the cycle of the effective angle of attack of the wing, and its return to zero is delayed due to the kinematics of wing folding (supplementary Fig.~\ref{sfig:eAoA}B), and the eAoA is high ($\sim \pm 50^\circ$) at the stroke reversals ($t/T = 0, 0.5$) and does not drop to zero until $t = 0.1$ and 0.7 respectively. As before, the quasi-steady analysis predicts zero lift when the eAoA is zero (as it should), while the measurements show that the lift has already dropped to zero ahead of the zero eAoA point due to the LEV separation.

At $t/T \sim 0.65$, during the upstroke of the high folding case, $\theta_o = 100$, the wings are almost perpendicular to the ground (Fig. \ref{fig:kineSnaps}),leading to a near zero prediction of $C_L$ from the quasi-steady model. The direct force measurement demonstrates a negative trough at this time, which may be due to the LEV on the dorsal side of the wing that generates unfavorable (negative) lift as it separates into the wake.  However, shortly after this, at $t/T \sim 0.8$ (stage iii in Fig.~\ref{fig:kineSnaps}), while the model still predicts negligible lift due to the vertical orientation of the wing surfaces (Fig~\ref{fig:kineSnaps}), the force measurements record a positive lift peak, also captured by the PIV measurements.  This lift augmentation arises from wing clapping(Fig.~\ref{fig:temporalAndScalingPIV}B), as a brief jet of air is squeezed downwards, similar to jetting in swimming jellyfish or squid \cite{Costello2020}.  This  momentary rise of $C_L$ due to wing clapping is not captured by the quasi-steady analysis, which cannot incorporate wing-wing interactions. The wing motors must provide extra energy to produce the jet and this additional power requirement is also absent from the quasi-steady model (Fig.~\ref{fig:temporalAndScalingPIV}A6, $t/T \sim 0.8$, see also snapshot iii in Fig.~\ref{fig:kineSnaps}).

\subsection{Aerodynamic force, power and efficiency of wing folding}

Birds and bats adjust wing folding amplitude with flight speed \cite{Hubel2016,Parslew2012,Riskin2012}. But among diverse robotic flyers, few studies have implemented wing folding \cite{send2012,Wissa2012,Qin2021,Chen2021,Qin2021}. Of these, Wissa et al. \cite{Wissa2012} employed passive mechanisms, while Send et al. \cite{send2012} implemented active wing folding with fixed amplitude and timing. Chen et al. \cite{Chen2021}  used a string-based approach to actuate variable-amplitude wing folding and found that wing folding helps reduce negative lift during upstroke in hovering condition. Through PIV measurements along the streamwise center plane of their robots, they demonstrated that the folded wings in upstroke are free of the impact of low pressure region under the wings, which otherwise would contribute to more negative lift.

Without wing folding ($\theta_o = 0$), the positive lift generated during the downstroke is balanced by the negative lift produced during the upstroke (Fig.~\ref{fig:cycAverPwrEco}A). For a given Strouhal number, $St$, as wing folding,$\theta_o$, increases, we observe two benefits.  First, handwing effective velocity increases because of the additional rotation with respect to the armwing, and this enhanced velocity increases the local effective angle of attack. As a result, both factors can contribute to greater lift (snapshot ii, Fig.~\ref{fig:temporalAndScalingPIV}A4 and Fig.~\ref{fig:kineSnaps}).  Second,  the effective wing surface area is reduced during upstroke, which alleviates negative lift, as has also been demonstrated in CFD simulations  \cite{Lang2022}. The additive effect of these factors results in net positive cycle-averaged lift, $C_L$, that increases with the folding amplitude, $\theta_o$, (Fig.~\ref{fig:cycAverPwrEco}A). 
    
We observe a qualitatively distinct phenomenon when wing folding amplitude $\theta_o$ exceeds a certain threshold  (see schematics in Fig.~\ref{fig:temporalAndScalingPIV}C).  The handwing lags the armwing, and when the armwing transitions from down- to upstroke, the handwing continues through its `downstroke' trajectory. The two wingtips move toward one another as they approach the midline, in a manner that can be described heuristically as ``scooping'' the air, producing a trough of negative lift (Fig.~\ref{fig:temporalAndScalingPIV}A4, between snapshots ii and iii). This phenomenon is also observed in small hovering birds \cite{Chang2011,Chang2013}.

As the handwings approach and ultimately clap (Fig.~\ref{fig:temporalAndScalingPIV}B), air is propelled dorsally and ventrally from the small gap between the wings (Fig.~\ref{fig:temporalAndScalingPIV}C). The net effect is a downward, ventral, momentum jet that produces positive lift, comparable in magnitude to that generated during downstroke (Fig.~\ref{fig:temporalAndScalingPIV}A4). In contrast, when Gouldian Finches clap, they position their wings to prevent dorsally-directed air movement \cite{Chang2013}).
    
The wingtips separate after the clap and are re-positioned prior to the next downstroke (snapshot iv, Fig.~\ref{fig:temporalAndScalingPIV}A4). With no modification of the movement, this preparatory movement could incur a substantial negative lift penalty, however, birds and bats adopt similar mitigation strategies. In both groups, the handwing supinates (rotates about the spanwise, long axis in the palm-upward direction) at the wrist and the wing retracts along the span by flexion at the elbow. The posture of the handwing may come close to a vertical plane before moving dorsally -- as it ``slices'' through air -- during the upstroke \cite{Hedrick2004,fan2021a,Hubel2010,Sekhar2018}. Upstroke kinematics of bats and birds reduce negative lift \cite{Chang2011,Windes2018,Hubel2010} and generate a small amount of thrust \cite{Hedenstrom2007,Muijres2014}. The wings of Flapperoo presented in this manuscript can neither twist nor retract, and the resulting unfavorable orientation of wing surfaces during rapid upstrokes leads to substantial negative lift. However, we did then designed, built and tested a Flapperoo with twisting capability, and found that twisting significantly decreased both power consumption, and negative lift during upstroke (accepted to present in IROS 2024, arXiv link \cite{Fan2024}).

In summary, two different mechanism to obtain optimal efficiency, characterized by power economy, are displayed here, and we may classify them as symmetry-breaking lift augmentation (small folding amplitude $\theta_o$) and appendage-based jet propulsion (extreme folding amplitude $\theta_o$). At low speed or high $St$ (Fig.~\ref{fig:cycAverPwrEco}B), it takes extra effort to actuate large wing folding to effect air jet for extra lift, thus the symmetry-breaking strategy (moderate folding amplitude $\theta_o \sim 30^\circ$ ) is more cost effective, in terms of $C_L/C_P$; on the other hand, for fast flight speed or low $St$, while symmetry-breaking strategy is still rewarding, jet propulsion with large folding angle becomes much more efficient (Fig.~\ref{fig:PwrEcoVsFA}). It is remarkable to note that flapping wing flight has both strategies at its disposal.

\subsection{Similarities and differences with ``clap-and-fling''}

The characteristic ``clap-and-fling'' mechanism of insects and slowly flying birds \cite{Weis-Fogh1973,Lehmann2015,Crandell2015} relies on the two wings coming together during the wingbeat cycle, but it differs fundamentally from the ventral clapping motion considered here. In clap-and-fling, the `fling' part of the motion generates thrust and enhances lift by augmenting circulation around the wing during early downstroke on the dorsal side \cite{Lehmann2015}. The ventral clapping mechanism that we describe here generates lift during upstroke by producing a directed jet of air. 

In both clap-and-fling and ventral wing clapping, however, negative lift is produced as a result of an upward jet, which can persist before and during the clapping ($C_L$ of case $\theta_o = 100$, Fig.~\ref{fig:temporalAndScalingPIV}A4) \cite{Chang2013,Lehmann2015}, Fig.~\ref{fig:temporalAndScalingPIV}B). 
The magnitude of net lift increase for insects, due to clap-and-fling, is small (Fig.~4 in \cite{Lehmann2015}), whereas we observe a more substantial boost, with a peak in the clapping-derived lift force comparable to that generated during the downstroke.  This distinction might be due to differences in the swept area. In clap-and-fling, the wings simultaneously rotate and translate (panel M-P in Fig.~7 in \cite{Lehmann2015}) so that the leading edge of the wings meet first, subsequently "squeezing" air out at the trailing edge. However, the volume of air trapped during this motion is much less than that the corresponding volume of air trapped by Flapperoo, where the armwings never come close and the handwings are on opposite sides of the relatively large body.

\subsection{Guidelines for flapping wing robots}

The potential benefits of ventral wing clapping, as demonstrated here, may inform future design of flapping wing robots with high payload or endurance requirements. Wing folding in birds and bats is implemented primarily by flexion and extension at the wrist joint \cite{Pennycuick2008,Norberg1990}. 

\textcolor{blue}{Even though a passively wrist joint with well-tuned stiffness to a specific speed may achieve beneficial folding angles, the flight efficiency may not adapt to the speed changes.}For example, the cycle-averaged power consumption of one passively deforming, bird-inspired ornithopter, Park Hawk \cite{kinkade2001}, was approximately $60 W$, which is considerably higher than the $20 W$ power requirement reported for SmartBird, a seagull-inspired robot that performs active wing folding and twist \cite{send2012, Wissa2012}, even though SmartBird has a greater wingspan and weighs more than Park Hawk. 
Here, we find that the cycle-averaged lift increases monotonically with wing folding angle across $St = 0.2 - 0.4$ (Fig.~\ref{fig:cycAverPwrEco}A). This translates directly to the capacity to carry larger payloads, and suggests that the energetic penalty for actuating the wing folding motion provides an effective compromise between generating lift and energetic cost (Fig.~\ref{fig:cycAverPwrEco}B). 

The controlled, selective variation in wing folding angles with flight speed by birds and bats \cite{Hubel2016,Fan2022,Parslew2012}, suggests that a flexible capability to modulate lift generation relative to energetic cost confers performance benefits. Adjustable folding angle could be incorporated in future designs for flapping wing robots to enhance their performance. For example, when agility is important, such as when navigating through a tight turn, an instantaneous boost of lift conferred by wing clapping might provide additional benefits with little impact on overall energetics due to the brief and temporary increase in power consumption. Moreover, ventral clapping is observed during fast flight ($U = 7.4$ m/s or $St$ \textasciitilde{} $0.2$ \cite{Fan2022}) in \Cb, where larger thrust is likely required to overcome high drag. 

\textcolor{blue}{Indeed, the next generation of Flapperoo is also able to perform wing twists \cite{Fan2024}, and demonstrates the air jet due to ventral wing clapping, as discussed in this paper,} can be directed wing angling at the moment of clap, where the leading edge almost touch but the trailing edge positioned slightly apart. This angled clapping produce streamwise propulsion (thrust) in addition to lift. This novel appendage-based propulsion offers many great opportunities to further probe into the optimal wing kinematics involving highly unsteady vortex dynamics, and also offers new control authority in both lift and thrust directions.



\bibliography{R12}
\bibliographystyle{unsrtnat}

\subsection*{Author Affiliations}


\newpage
\onecolumn
\setcounter{page}{1}
\setcounter{section}{0}
\setcounter{equation}{0}
\setcounter{figure}{0}
\def\theequation{S\arabic{equation}}
\renewcommand{\thetable}{S\arabic{table}}
\renewcommand{\thefigure}{S\arabic{figure}}

\section*{Supplementary Materials}

\begin{figure}[ht]
\centering
\includegraphics[width=3in]{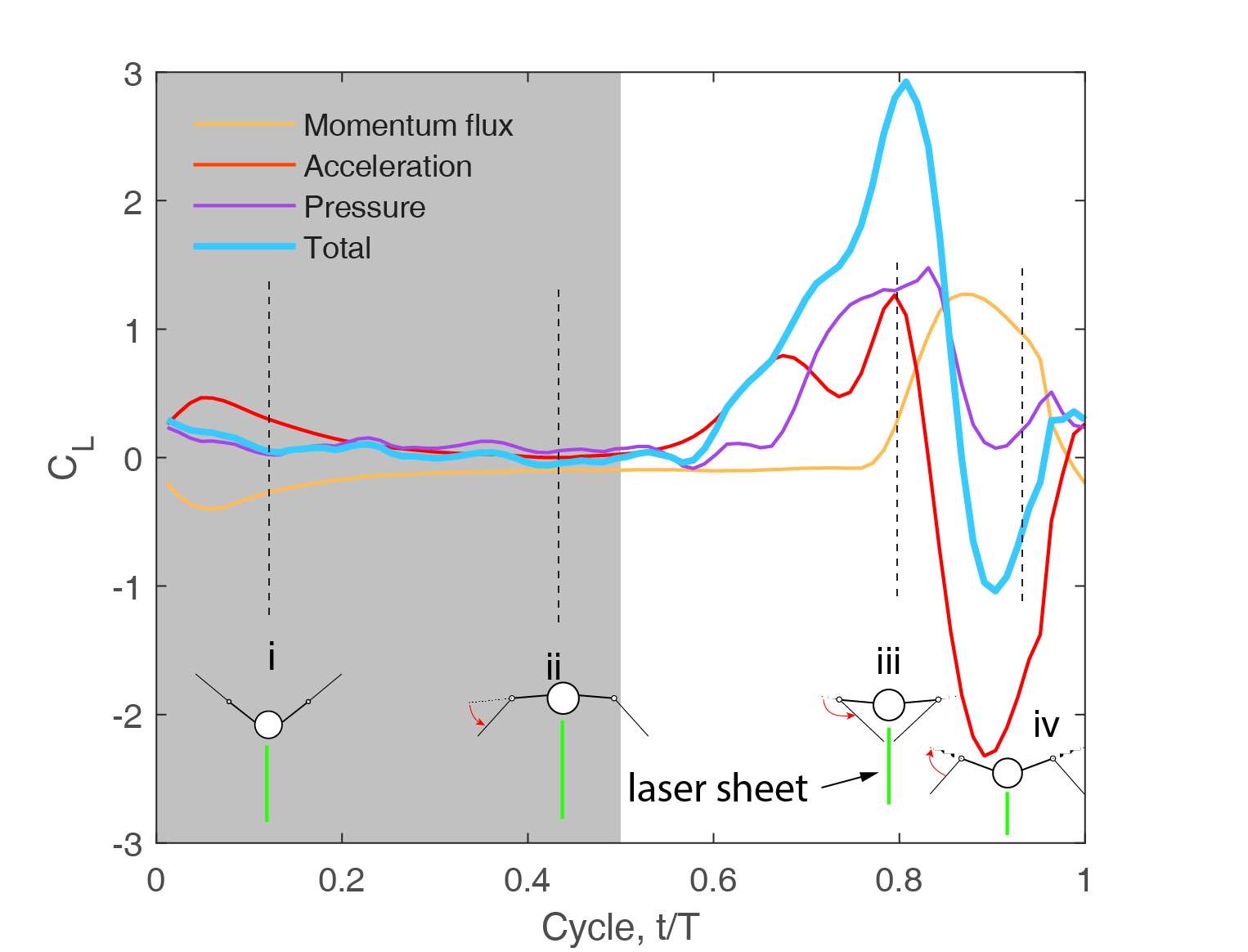}
\caption{ (Supplementary) Time series of control volume analysis for $C_{L}$ and its breakdown as a function of cycle for $St = 0.2$, and fold amplitude $\theta_o = 100$. The results are based on phased-averaged flow fields from PIV experiments for $N = 40$ cycles. Insets show the position of laser sheet with respect to the robotic platform. The measurement captures relevant flow features when the wings are near the laser plane. Near the moment of wing clapping (snapshot iii), the lift contribution of the jet, $C_{L,clap}$, is mostly unsteady pressure and the acceleration of fluid particles, followed by a delayed contribution from convective momentum flux.}
\label{fig:temporalJetComp}
\end{figure}

From the control volume (CV) analysis, we can identify three primary components of lift: the downward acceleration of the fluid inside the control volume, the forces due to the change in  pressure between the top and bottom boundaries, and the vertical momentum convected out of the control volume (Fig.~\ref{afig:CV setup} and Fig.~\ref{fig:temporalJetComp}). Pressure variation and the acceleration of the air inside the control volume are the two main contributors to the clap-generated jet at the instant of the clap, while the momentum flux lags in phase because of the time required for the fluid to convect to the downstream boundary of the control volume (surface 2 in Fig.~\ref{afig:CV setup}). 

\begin{figure}[ht]
\centering
\includegraphics[width=2.5in]{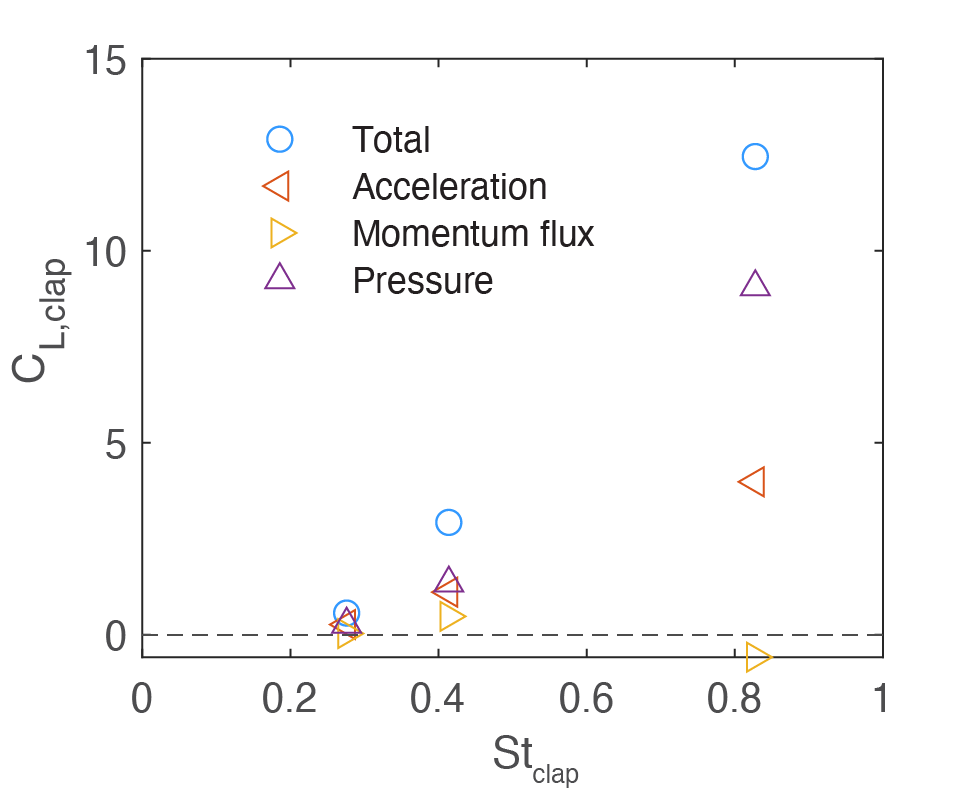}
\caption{ (Supplementary) $C_{L,clap}$ and its components as a function of Strouhal number $St_{clap}$ for wing fold amplitude $\theta_o = 100$. The results are based on phased-averaged flow field from PIV experiment for $N = 40$ cycles.}
\label{fig:jetCompVSSt}
\end{figure}

\begin{figure}
\centering
\includegraphics[width=3.5in]{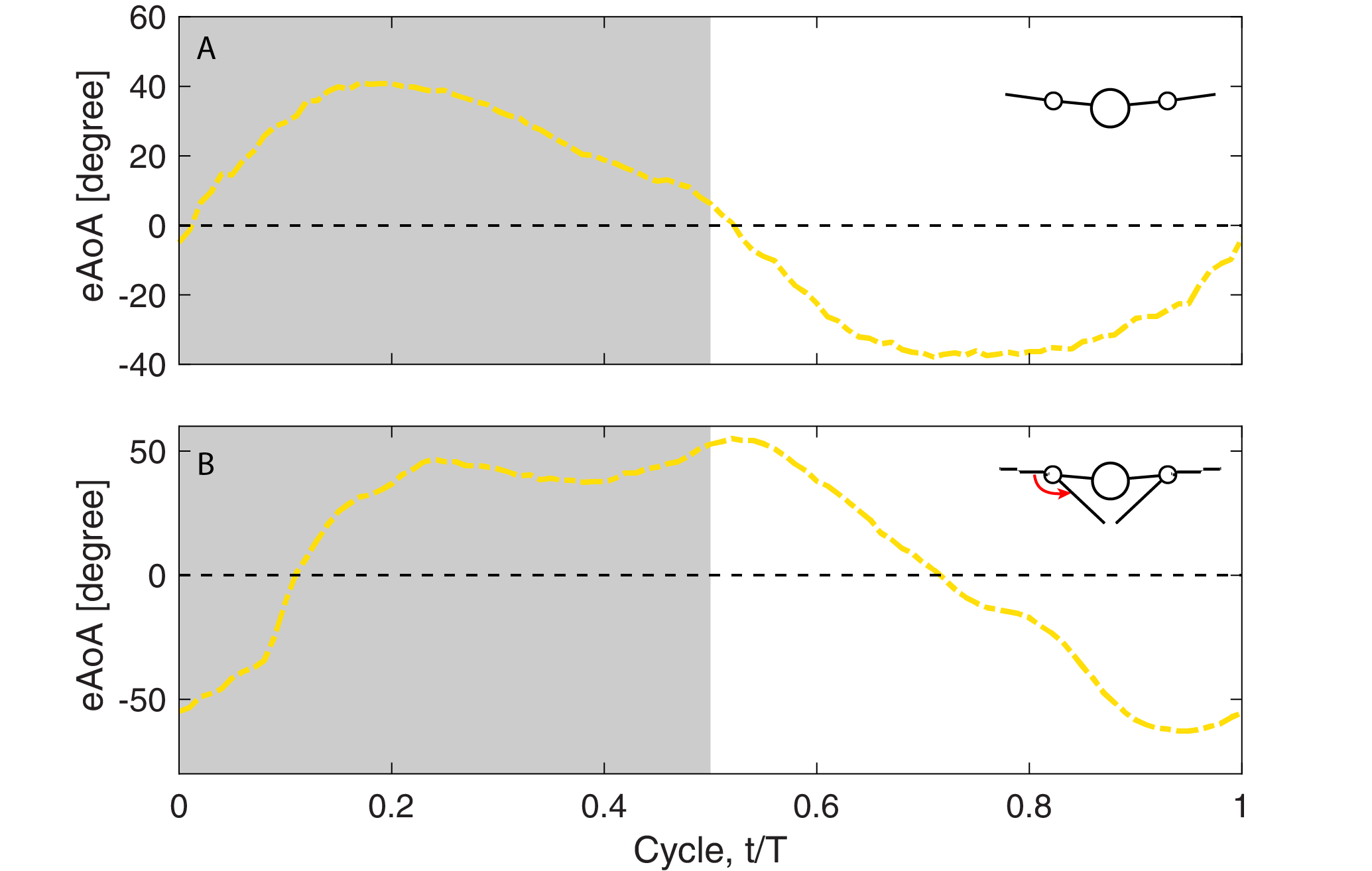}
\caption{ (Supplementary) Effective angle of attack evolution over the wingbeat cycle at the $3/4$ span section for folding amplitude (A) $\theta_o = 0$ and (B) $\theta_o = 100$. }
\label{sfig:eAoA}
\end{figure}

$C_{L,clap}$ and its components, the acceleration of the fluid parcels and the net pressure, vary as a function of $St_{clap}$, and the momentum flux stays around zero (Fig.~\ref{fig:jetCompVSSt}). Over the $St_{clap}$ studied here, the pressure difference between the top and bottom control surfaces makes the largest contribution to $C_{L,clap}$, followed by the fluid inertia. The convective effect is negligible around the moment of wing clapping because the vertical momentum flux has not yet been transported by the freestream to the downstream surface of the control volume.

\begin{figure}
\centering
\includegraphics[width=3.5in]{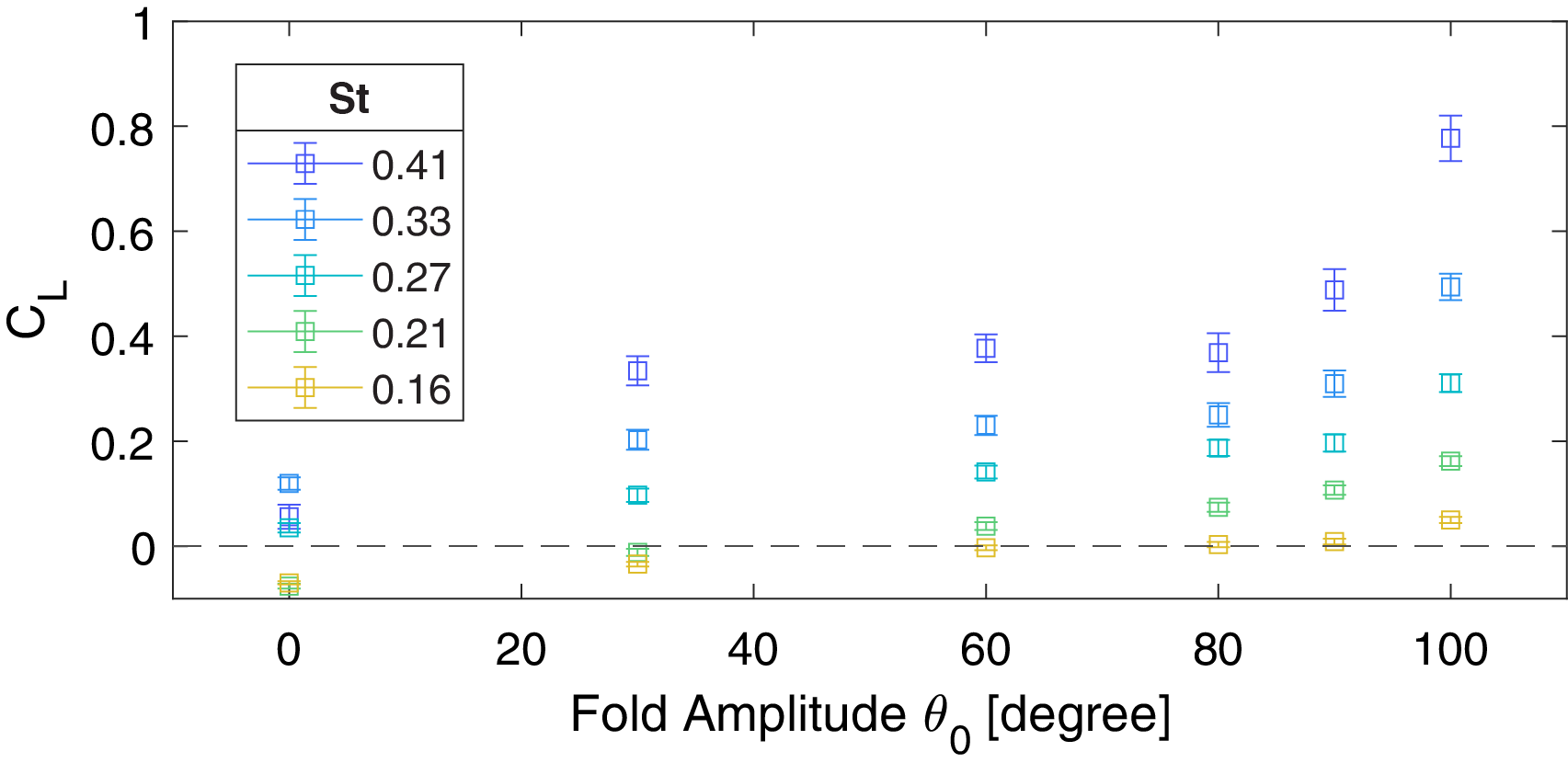}
\caption{ (Supplementary) Cycle-averaged lift coefficient $C_L$ as a function of fold amplitude $\theta_o$.}
\label{sfig:ClvsFold}
\end{figure}


\begin{figure}
\centering
\includegraphics[width=3.5in]{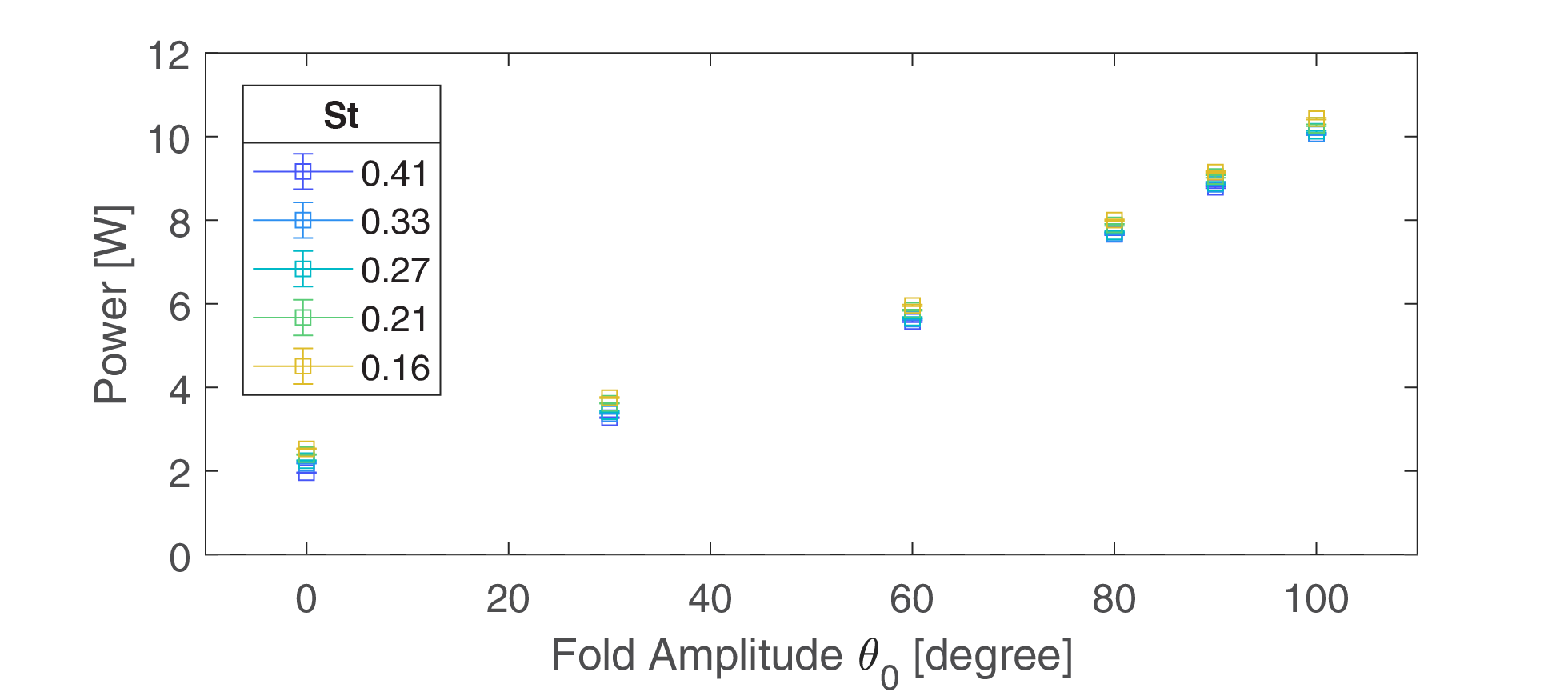}
\caption{ (Supplementary) Cycle-averaged power consumption as a function of fold angle $\theta_o$.}
\label{sfig:PwrvsFold}
\end{figure}

\begin{figure}
\centering
\includegraphics[width=3.5in]{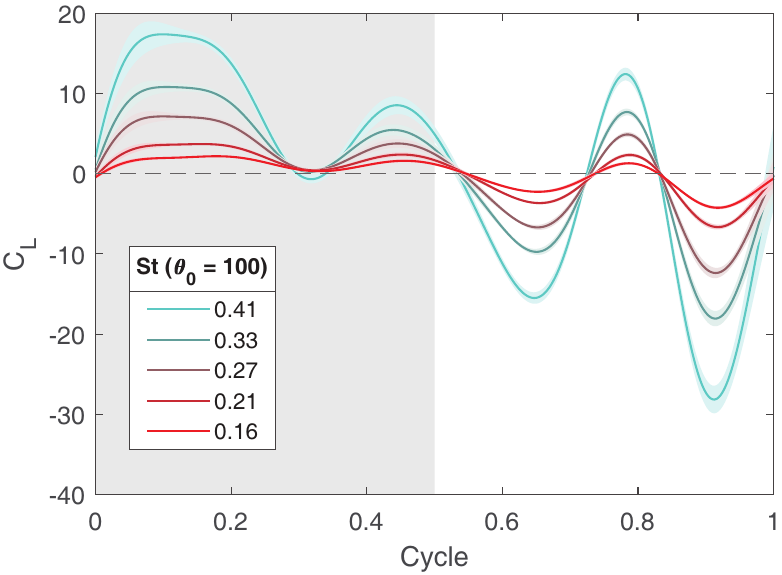}
\caption{ (Supplementary) Lift coefficient $C_L$ as a function of cycle normalized time for different Strouhal numbers. The wings would clap for cases presented here or folding angle $\theta_o = 100$. }
\label{sfig:AllClvsTime}
\end{figure}

\begin{figure}
\centering
\includegraphics[width=5in]{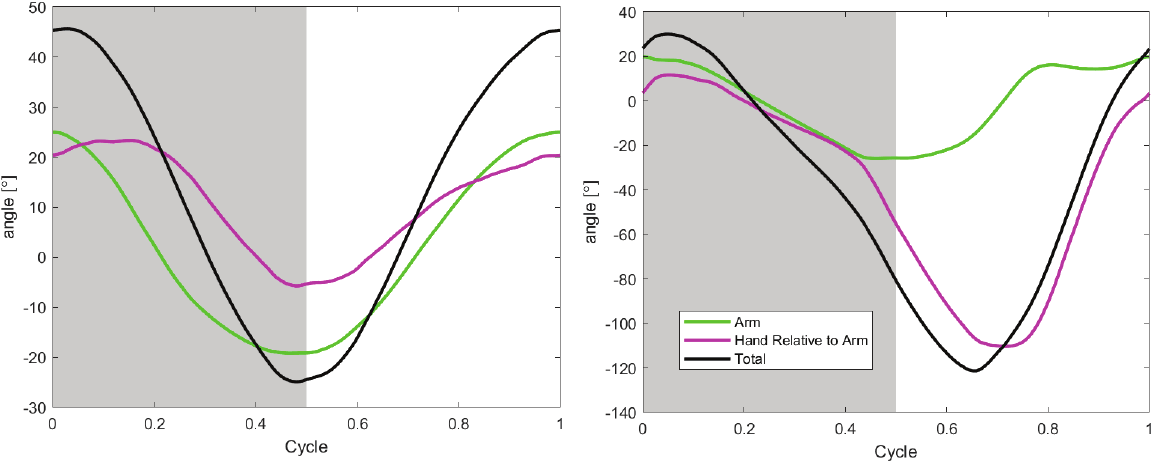}
\caption{ (Supplementary) Recorded wing kinematics for (A) $\theta_0 = 0$, and (B) $\theta_0 = 100$. Shaded area refers to downstroke. Green line is armwing flapping angle, purple is handwing relative to armwing angles, and black line is the total handwing angle relative to mid-stroke plane.}
\label{sfig:hdMid}
\end{figure}

\begin{figure}
\centering
\caption{ (Supplementary video f3U0S50Nbin83cy40.gif) Time resolved PIV streamwise plane. Part of the video is overlaid with wing imaging. }
\label{sfig}
\end{figure}

\newpage



\section*{Control volume analysis derivation details}

 \begin{figure*}
 \includegraphics[width=7in]{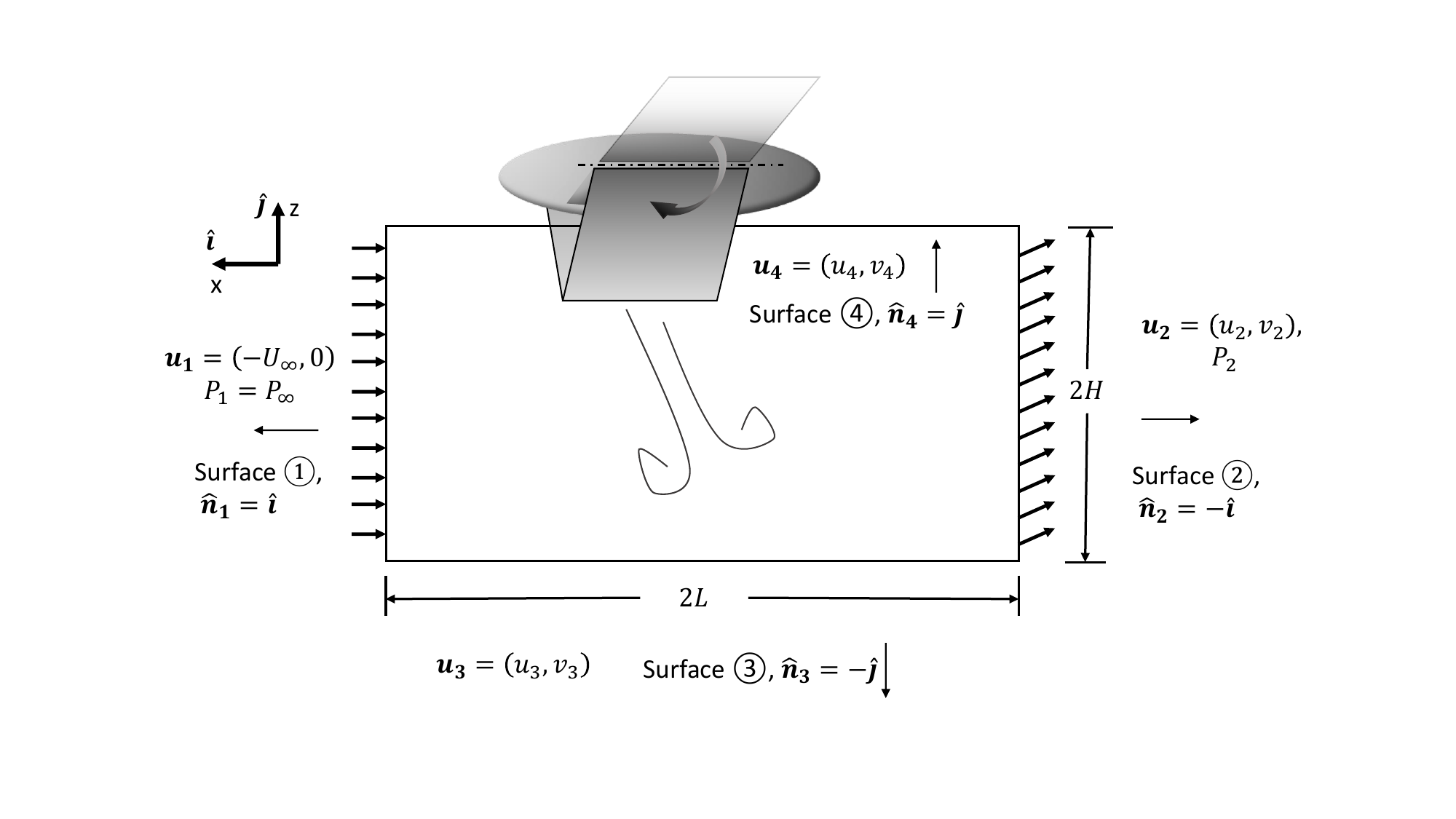}
 \caption{Derivation of control volume analyses for lift estimation}
 \label{afig:CV setup}
 \end{figure*}

In order to quantitatively understand which term(s) (unsteadiness, momentum flux or pressure difference) dominate(s) the physics right around wing clapping, we utilize the flow field data from PIV experiment (Fig.~\ref{fig:temporalAndScalingPIV}B) to reconstruct the lift force.

Specifically, the aerodynamic forces that result from wing clapping, $\vec{L}$, may be obtained from the Navier-Stokes (N-S) equation for a control volume (C.V.) analysis approach. 
Here, as outlined in Fig.~\ref{afig:CV setup}, $x-$ axis points to the direction of flight, and $z-$ axis points to vertically upward. Freestream $U_\infty$ is in the negative $x$ direction.
The viscous effect is negligible as the Reynolds number $Re$ is around $10^5$ for our Flapparoo ventral clapping experiment, the N-S equation is then given as: 

\begin{equation}\label{eq:NS CV}
\underbrace{\frac{\partial}{\partial t} \int_{C . V .} \rho \vec{v} d\forall}_{\mathrm{I}}+
\underbrace{\int_{C . S .} \rho \vec{v}(\vec{v} \cdot \vec{n}) d A}_{\mathrm{II}}
=\underbrace{\int_{C . S .}-p \cdot \vec{n} d A}_{\text {III }}
+ \vec{L}
\end{equation}

Here, the control volume (C.V.) encapsulates the space where wing would clap, with control surface (C.S.) on the four boundaries, indexed by 1,2,3 and 4 . 

The first term (term I in eq.~\ref{eq:NS CV}) describes the unsteadiness (acceleration and deceleration of fluid parcels) within the control volume, and if we look at the z-direction ($\hat{\mathbf{j}}$), we have: 

\begin{equation}
\hat{\boldsymbol{\jmath}} \cdot \frac{\partial}{\partial t} \int_{C.V.} \rho \vec{v} d \forall
\end{equation}

Since the control volume is fixed in space, we may absorb the differentiation with respect to time under the integral sign, and extract just the $z-$axis component of the velocity vector. Also notice the air density $\rho$ is a constant, and can be pulled outside of the integral as:

\begin{equation}\label{eq:I-unsteady}
\rho \int_{C.V.} \frac{\partial v}{\partial t} d \forall 
= \rho \int_{C.V.} a_z d \forall,
\end{equation}

Here, we use $a_z$ to represent the acceleration of fluid particles. Since the velocity field from PIV experiment can be noisy and subject to fluctuations due to turbulence and other flow instabilities, phase averaging is an appropriate technique to apply, which involves collecting multiple records of the same flow field, and averaging across these the velocity fields records at each spatial location to get a time-dependent mean value, and a zero-mean fluctuation. Such a technique reduces the noise in the velocity field, and provides a more accurate representation of the dominant flow field. 

Specifically in our study, we perform phase-averaging to all the physical quantities obtained, \textit{e.g.}, for $a_z$ we may decompose it as sum of a phase-averaged mean ($A_z$), and a fluctuation term ($a_z^\prime$) as $a_z = A_z + a_z^\prime$. By definition, the fluctuation term averages to zero across all records, or,

\begin{align}
& \quad \rho \int_{C.V.} (a_z)_\text{phase-averaging} d \forall \\
&= \rho \int_{C.V.} (A_z + a_z^\prime)_\text{phase-averaging} d \forall  \\
&= \rho \int_{C.V.} A_z d \forall, 
\end{align}

We normalize with $\rho U_\infty^2c/2 $ to get non-dimensional force per unit length into the paper, where $U_\infty$ and $c$ are freestream and chord length respectively. The z-direction non-dimensional unsteadiness term is thus

\begin{equation}
\frac{2}{c} \frac{1}{U_{\infty}^2} \int_{C . V .} A_z d \forall
\end{equation}

Similarly, for the second flux term (term II in eq.~\ref{eq:NS CV}) along the z-direction ($\hat{\mathbf{j}}$), we have:

\begin{equation}
\begin{aligned}
& \hat{\boldsymbol{\jmath}} \cdot \int_{C . S .} \rho \vec{v}(\vec{v} \cdot \vec{n}) d A \\
& =\int_1 \rho 0\left(-U_{\infty}\right) d z+\int_2 \rho v_2\left(-u_2\right) d z+\int_3 \rho v_3\left(-v_3\right) d x+\int_4 \rho v_4\left(v_4\right) d x \\
& =-\int_2 \rho v_2\left(u_2\right) d z - \int_3 \rho v_3^2 d x + \int_4 \rho v_4^2 d x,
\end{aligned}
\end{equation}

where the subscripts 1,2,3 and 4 index control surfaces in Fig.~\ref{afig:CV setup}. Now, we decompose each velocity into a phase-averaged mean and the associated fluctuation, \textit{i.e.}, $v_2=V_2+v_2^{\prime}$, $v_3=V_3+v_3^{\prime}$, $v_4=V_4+v_4^{\prime}$ and $u_2=U_2+u_2^{\prime}$, and insert these expression into the above equation:

\begin{equation}
\begin{aligned}
& -\int_2 \rho\left(V_2+v_2^{\prime}\right)\left(U_2+u_2^{\prime}\right) d z-\int_3 \rho\left(V_3+v_3^{\prime}\right)^2 d x+\int_4 \rho\left(V_4+v_4^{\prime}\right)^2 d x \\
& =-\int_2 \rho\left(V_2 U_2+V_2 u_2^{\prime}+v_2^{\prime} U_2+v_2^{\prime} u_2^{\prime}\right) d z-\int_3 \rho\left(V_3^2+2 V_3 v_3^{\prime}+v_3^{\prime 2}\right) d x \\
& +\int_4 \rho\left(V_4^2+2 v_4^{\prime} V_4+v_4^{\prime 2}\right) d x
\end{aligned}
\end{equation}

Note that the primed ($\prime$) terms will be phase-averaged to zero, thus we are left with: 

\begin{equation}
-\int_2 \rho\left(V_2 U_2+v_2^{\prime} u_2^{\prime}\right) d z-\int_3 \rho\left(V_3^2+v_3^{\prime 2}\right) d x+\int_4 \rho\left(V_4^2+{v_4^{\prime}}^2\right) d x
\end{equation}

We then non-dimensionalize the above expression using $\rho U_\infty^2 c /2$ to get non-dimensional force per unit length into the paper, and arrive:

\begin{equation}
\frac{2}{c} \frac{1}{U_{\infty}^2}\left(-\int_2\left(V_2 U_2+v_2^{\prime} u_2^{\prime}\right) d z-\int_3\left(V_3^2+v_3^{\prime 2}\right) d x+\int_4\left(V_4^2+v_4^{\prime 2}\right) d x\right)
\end{equation}

To evaluate the last unsteady pressure term (III) in eq.~\ref{eq:NS CV}, we follow similar steps:

\begin{equation}
\begin{aligned}\label{eq:unsteadyPressureIncomplete}
& \hat{\boldsymbol{\jmath}} \cdot \int_{C . S .}-p \cdot \vec{n} d A \\
& =\left(\int_1-p_{\infty} \cdot \hat{\boldsymbol{j}} \cdot \vec{n}_{1} d z+\int_2-p_2 \cdot \hat{\boldsymbol{\jmath}} \cdot \vec{n}_2 d z+\int_3-p_3 \cdot \hat{\boldsymbol{\jmath}} \cdot \vec{n}_3 d x\right. \\
& \left.+\int_{4}-p_4 \cdot \hat{\boldsymbol{\jmath}} \cdot \vec{n}_4 d x\right) \\
& =\left(\int_3-p_3 \cdot(-1) d x+\int_4-p_4 \cdot d x\right)=\int_{-L}^L\left(p_3-p_4\right) d x
\end{aligned}
\end{equation}

However, since we did not directly measure pressure in our experiment, we need the $x-$direction momentum equation to obtain pressure through velocities. Now,

\begin{equation}
\frac{\partial u}{\partial t}+u \frac{\partial u}{\partial x}+v \frac{\partial u}{\partial z}=-\frac{1}{\rho} \frac{\partial p}{\partial x}
\end{equation}

Similarly as before, we decompose each velocity and pressure into a phase-averaged mean and a zero-mean fluctuation, such as $u=U+u', v=V+v' \text{and } p = P + p'$, and insert back into the equation above, which yields:

\begin{equation}
\begin{aligned}
& \frac{\partial\left(U+u^{\prime}\right)}{\partial t}+\left(U+u^{\prime}\right) \frac{\partial\left(U+u^{\prime}\right)}{\partial x}+\left(V+v^{\prime}\right) \frac{\partial\left(U+u^{\prime}\right)}{\partial z} \\
& =-\frac{1}{\rho} \frac{\partial\left(P+p^{\prime}\right)}{\partial x},
\end{aligned}
\end{equation}

Expanding each term, we have:

\begin{equation}
\begin{aligned}
& \frac{\partial U}{\partial t}+\frac{\partial u^{\prime}}{\partial t}+U \frac{\partial U}{\partial x}+U \frac{\partial u^{\prime}}{\partial x}+u^{\prime} \frac{\partial U}{\partial x}+u^{\prime} \frac{\partial u^{\prime}}{\partial x}+V \frac{\partial U}{\partial x}+V \frac{\partial u^{\prime}}{\partial x}+v^{\prime} \frac{\partial U}{\partial x}+ \\
& v^{\prime} \frac{\partial u^{\prime}}{\partial x}=-\frac{1}{\rho}\left(\frac{\partial P}{\partial x}+\frac{\partial p \prime}{\partial x}\right) ;
\end{aligned}
\end{equation}

After phase-averaging, the survived terms are: 
\begin{equation}
\frac{\partial U}{\partial t}+U \frac{\partial U}{\partial x}+u^{\prime} \frac{\partial u^{\prime}}{\partial x}+V \frac{\partial U}{\partial x} + v^{\prime} \frac{\partial u^{\prime}}{\partial x}=-\frac{1}{\rho} \frac{\partial P}{\partial x}
\end{equation}

Then, integrate the phase-averaged pressure $P$ along $x-$axis from left boundary L to any location x (Fig.~\ref{afig:CV setup}), we have:

\begin{equation}
\begin{aligned}\label{eq:pressure}
\int_L^x\left(\frac{\partial P}{\partial x}\right) d x= & -\rho \int_L^x\left(\frac{\partial U}{\partial t}+U \frac{\partial U}{\partial x}+u^{\prime} \frac{\partial u^{\prime}}{\partial x}+V \frac{\partial U}{\partial x}+v^{\prime} \frac{\partial u^{\prime}}{\partial x}\right) d x \\
P(x, z)-P(x= & L, y)=-\rho\left(\int_L^x \frac{\partial U}{\partial t} d x+\frac{1}{2} \int_L^x \frac{\partial U^2}{\partial x} d x\right. \\
& \left.+\frac{1}{2} \int_L^x \frac{\partial u^{\prime 2}}{\partial x} d x+\int_L^x V \frac{\partial U}{\partial x} d x+\int_L^x v^{\prime} \frac{\partial u^{\prime}}{\partial x} d x\right) \\
P(x, z)-P_{\infty}= & -\rho\left(\int_L^x A_x d x+\frac{1}{2}\left(U^2-U_{\infty}^2\right)\right. \\
& \left.+\frac{1}{2} u^{\prime 2}+\int_L^x \bar{v} \frac{\partial \bar{u}}{\partial x} d x+\int_L^x v^{\prime} \frac{\partial u^{\prime}}{\partial x} d x\right)
\end{aligned}
\end{equation}

Evaluate eq.~\ref{eq:pressure} at bottom surface 3 ($z = -H$) and top surface 4 ($z = H$) in Fig.~\ref{afig:CV setup}, we have
\begin{equation} \label{eqPressuresurf3}
\begin{aligned}
P(x, -H)-P_{\infty}= & -\rho\left(\int_L^x A_{x, 3} d x+\frac{1}{2}\left(U_3{ }^2-U_{\infty}^2\right)+\frac{1}{2} u_3^{\prime 2}\right. \\
& \left.+\int_L^x V_3 \frac{\partial U_3}{\partial x} d x+\int_L^x v_3^{\prime} \frac{\partial u_3^{\prime}}{\partial x} d x\right) \\
\end{aligned}
\end{equation}

\begin{equation} \label{eqPressuresurf4}
\begin{aligned}
P(x,H)-P_{\infty}= & -\rho\left(\int_L^x A_{x, 4} d x+\frac{1}{2}\left(U_4{ }^2-U_{\infty}^2\right)+\frac{1}{2} u_4^{\prime 2}+\right. \\
& \left.\int_L^x V_4 \frac{\partial U_4}{\partial x} d x+\int_L^x v_4^{\prime} \frac{\partial u_4^{\prime}}{\partial x} d x\right)
\end{aligned} 
\end{equation}

Finally, by subtracting eq.~\ref{eqPressuresurf4} from eq.~\ref{eqPressuresurf3}, and normalizing by $\rho U_\infty^2 c/2$, we arrive at: 

\begin{equation}
\begin{aligned}
& P(x, -H)-P(x,H)=P_3-P_4 \\
& =\frac{2}{c U_{\infty}^2} \int_L^x\left(A_{x, 4}-A_{x, 3}\right) d x+\frac{1}{2}\left(U_4{ }^2-U_3{ }^2\right)+\frac{1}{2}\left({u_4^{\prime}}^2-{u_3^{\prime}}^2\right) \\
& +\int_L^x\left(V_4 \frac{\partial U_4}{\partial x}-V_3 \frac{\partial U_3}{\partial x}\right)d x+\int_L^x\left(v_4^{\prime} \frac{\partial u_4^{\prime}}{\partial x}-v_3^{\prime} \frac{\partial u_3^{\prime}}{\partial x}\right) d x
\end{aligned}
\end{equation}

Note, the pressure difference is now represented purely as velocities and its derivatives. Plugging the above expression back into eq.\ref{eq:unsteadyPressureIncomplete}, then the unsteady pressure becomes:

\begin{equation}
\frac{2}{c U_{\infty}^2} \int_{-L}^L\left(\begin{array}{c}
\int_L^x\left(A_{x, 4}-A_{x, 3}\right) d x+\frac{1}{2}\left(U_4^2-U_3^2\right)+\frac{1}{2}\left(u_4^{\prime 2}-u_3^{\prime 2}\right) \\
+\int_L^x\left(V_4 \frac{\partial U_4}{\partial x}-V_3 \frac{\partial U_3}{\partial x}+v_4^{\prime} \frac{\partial u_4^{\prime}}{\partial x}-v_3^{\prime} \frac{\partial u_3^{\prime}}{\partial x}\right) d x
\end{array}\right) d x
\end{equation}

Finally, the lift coefficient $C_L$ is modeled as: 

\begin{equation}
\begin{aligned}
& C_L=\underbrace{\frac{2}{c} \frac{1}{U_{\infty}^2} \int_{\text {C.V. }} A_z d \forall}_{\text {Unsteadiness }} \\
& \underbrace{+\frac{2}{c} \frac{1}{U_{\infty}^2}\left(\int_2\left(V_2 U_2+v_2^{\prime} u_2^{\prime}\right) d y+\int_3\left(V_3^2+v_3^{\prime 2}\right) d x-\int_4\left(V_4^2+v_4^{\prime 2}\right) d x\right.}_{\text { Momentum flux }}) \\
& \underbrace{+\frac{2}{c U_{\infty}^2} \int_{-L}^L\left(\begin{array}{c}
\int_L^x\left(A_{x, 4}-A_{x, 3}\right) d x+\frac{1}{2}\left(U_4^2-U_3^2\right)+\frac{1}{2}\left({u_4^{\prime}}^2-u_3^{\prime 2}\right) \\
+\int_L^x\left(V_4 \frac{\partial U_4}{\partial x}-V_3 \frac{\partial U_3}{\partial x}+v_4^{\prime} \frac{\partial u_4^{\prime}}{\partial x}-v_3^{\prime} \frac{\partial u_3^{\prime}}{\partial x}\right) d x
\end{array}\right) d x}_{\text { Pressure }} \\
&
\end{aligned}
\end{equation}

\end{document}